\documentclass{article}

\usepackage{PRIMEarxiv}

\usepackage[utf8]{inputenc} % allow utf-8 input
\usepackage[T1]{fontenc}    % use 8-bit T1 fonts
\usepackage{hyperref}       % hyperlinks
\usepackage{url}            % simple URL typesetting
\usepackage{booktabs}       % professional-quality tables
\usepackage{amsfonts}       % blackboard math symbols
\usepackage{nicefrac}       % compact symbols for 1/2, etc.
\usepackage{microtype}      % microtypography
\usepackage{lipsum}
\usepackage{fancyhdr}       % header
\usepackage{graphicx}       % graphics

\graphicspath{{media/}}     % organize your images and other figures under media/ folder

\title{Flexible mm-Wave Frequency and High-Speed Arbitrary IQ Signal Synthesis by a Photonic System on Chip
}

\author{
  Bowen Zhu\textdagger\\
  Pengcheng Laboratory\\Shenzhen, 518055, China\\
  \texttt{ } \\
  \And
  Tao Zhu\textdagger\\
  Pengcheng Laboratory\\School of Integrated Circuits\\ Harbin Institute of Technology(Shenzhen)\\
  Shenzhen, 518055, China\\
  \texttt{ } \\
  \And 
  Yazhi Pi \\
  Pengcheng Laboratory\\Shenzhen, 518055, China\\
  \texttt{ } \\
  \And 
  Chunyang Ma \\
  Pengcheng Laboratory\\Shenzhen, 518055, China\\
  \texttt{ } \\
  \And 
  Xiaochuan Xu \\
  School of Integrated Circuits\\ Harbin Institute of Technology(Shenzhen)\\Pengcheng Laboratory\\Shenzhen, 518055, China\\
  \texttt{ } \\
  \And 
  Zizheng Cao \\
  College of Integrated Circuits\\Zhejiang University\\Hangzhou, 311200, China\\
  \texttt{zcaozju@zju.edu.cn} \\
  \And 
  Lei Wang \\
  Pengcheng Laboratory\\Shenzhen, 518055, China\\
  \texttt{wangl07@pcl.ac.cn} \\
  \And 
  Shaohua Yu \\
  Pengcheng Laboratory\\Shenzhen, 518055, China\\
  \texttt{yush@cae.cn} \\
}

\begin{document}
\maketitle

\begin{abstract}
Photonics-assisted millimeter-wave bands and terahertz signal generation offer significant advantages over traditional electronic methods by leveraging the inherent benefits of optical components, including broad bandwidth, low power consumption, and minimal insertion loss. This work utilizes a silicon photonic chip in conjunction with a reconfigurable optical frequency comb to demonstrate the synthesis of signals in the millimeter-wave range. The implemented photonic system performs on-chip filtering and modulation, producing high-bandwidth single frequency, multi-frequency, and vector signals suitable for arbitrary IQ signal construction. These results highlight the flexible and reconfigurable capabilities of the proposed approach, providing new perspectives for applications in radio-over-fiber systems and beyond.
\end{abstract}

% keywords can be removed

\section{Introduction}

Driven by the rapid advancement of emerging applications such as intelligent terminals, high-definition streaming, the metaverse, and digital twins, the global wireless network throughput has exhibited exponential growth\cite{zhu2023ultra}. To accommodate this trend, the carrier frequencies in wireless communications have been progressively increased to meet the demand for larger bandwidths. At present, the low-band  millimeter-wave (mm-wave) has been deployed in fifth-generation (5G) mobile communication systems, while higher-frequency mm-wave and terahertz-wave (THz-wave) signals are expected to enable next-generation wireless networks by overcoming bandwidth limitations\cite{nagatsuma2009generating,li2019photonics,li2022photonics,dat2025photonics}. 

\begin{figure}
    \centering
    \includegraphics[width=0.7\linewidth, height=0.472\textheight]{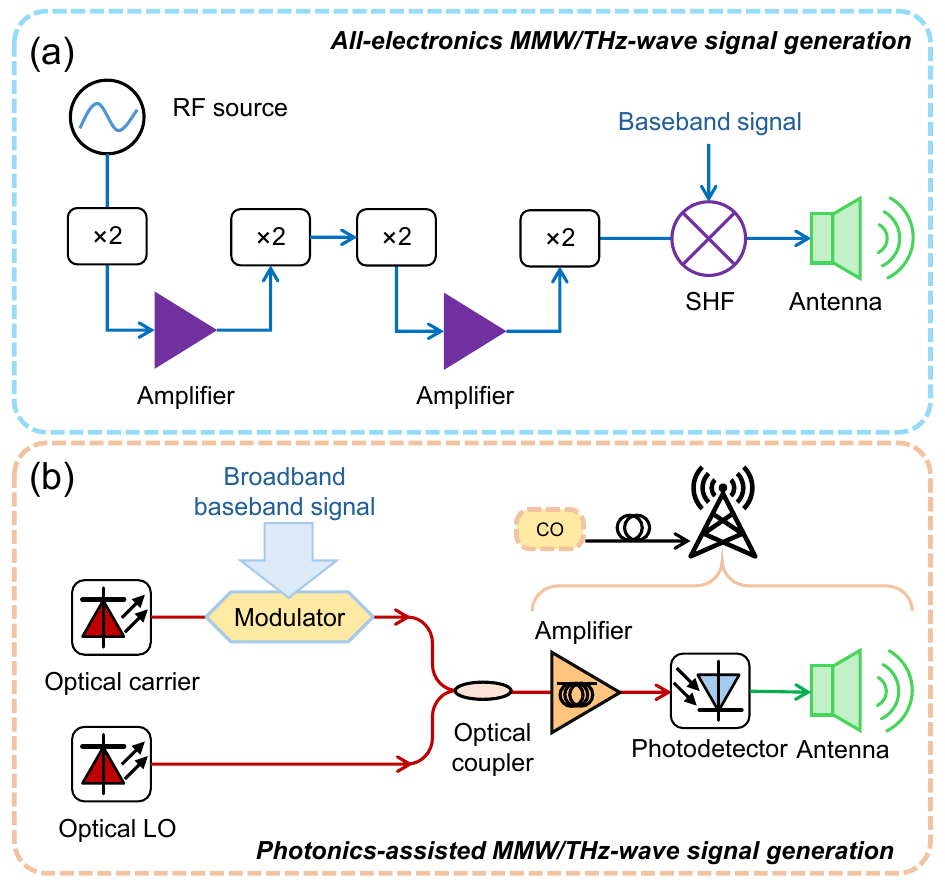}
    \caption{Schematic comparison between all-electronics MMW/THz-wave signal generation and Photonics-assisted approach.}
    \label{fig:schematic}
\end{figure}

Conventional purely electronic approaches for terahertz signal generation primarily rely on frequency multiplication and amplification of radio-frequency (RF) signals followed by mixing with baseband signals, as illustrated in Fig\ref{fig:schematic} (a). However, these methods face limitations due to electronic bandwidth constraints, and suffer from inflexible frequency tuning, poor compatibility with fiber-optic networks, and limited stability. Leveraging photonic technologies, RF signals can be generated by heterodyning two optical carriers through a high-speed photodiode (PD), allowing the system to exploit the inherent advantages of optical components-such as high frequency, broad bandwidth, and low transmission loss-thus breaking the electronic bandwidth bottleneck. One typical approach employs two independent lasers, as shown in figure\ref{fig:schematic} (b). Assume that the frequency of the optical carrier is $f_1$, and the frequency of the optical local oscillator (LO) is $f_2$. The modulated signal can be expressed as\cite{yu2021broadband}:

\begin{equation}
    E_s(t) = A_1·[I(t) + jQ(t)]·exp[j2\pi f_1t + j\theta _1(t)]
\end{equation}
Where $I(t) + jQ(t)$ represents the modulated baseband signal, $A_1$ represents the amplitude of the optical carrier. An optical coupler (OC) combines the modulated data carrier and LO, and the combined signal can be expressed as:

\begin{equation}
    E(t)=\frac{A_1·[I(t)+jQ(t)]·exp[j2\pi f_1t + j\theta _1(t)]+j·A_2·exp[j2\pi f_2t + j\theta _2(t)]}{\sqrt{2}} 
\end{equation}
The modulated signals can be transported seamlessly over extended distances leveraging optical fiber infrastructure, thereby overcoming the high propagation losses of mm-wave/THz-wave signals through air and enhancing the performance of hybrid systems. A photodiode (PD) performs heterodyne beating to generate high bandwidth signals. After being fed into the detector of square law detection, the electric signal current generated by two lights beat frequency can be expressed as:

\begin{equation}
    I(t) = RA_1^{2}[I^{2}(t)+Q^{2}(t)]+RA_2^{2}+2RA_1A_2[I(t)sin(2\pi \Delta f + \Delta \theta (t)) + Q(t)cos(2\pi \Delta f + \Delta \theta (t))]
\end{equation}
Where $R$ represents the responsivity of the photodetector, whose unit is A/W. $\Delta f$ is the difference between $f_1$ and $f_2$, $\Delta \theta$ is the difference between $\theta_1$ and $\theta_2$. The first two components in the current expression correspond to DC terms, while the third term carries the desired terahertz waveform. By tuning the frequency detuning between the two lasers to the terahertz band, heterodyne beating in a photodetector enables the generation of terahertz signals.

This approach has been successfully applied in radio-over-fiber (RoF) systems, demonstrating its practical viability for mobile fronthaul and high-frequency signal distribution\cite{lim2020radio,novak2015radio,o1994fibre,lima1995compact}. However, the heterodyne-generated signal in this scheme suffers from frequency offset and phase noise, which must be compensated using complex digital signal processing (DSP) algorithms for frequency and phase recovery\cite{zhu202240,zhu2022photonics,li2023photonics}. Alternatively, an optical frequency comb (OFC) derived from a single laser source can be utilized, where two desired optical carriers are selected via optical filtering and heterodyned to produce an RF signal with significantly improved frequency stability\cite{jia20180,lim2019evolution}.

Although photonics-assisted mm-wave and THz-wave systems have demonstrated the potential to achieve ultra-high-speed wireless transmission beyond 100 Gbps, most system-level implementations still rely on bulky discrete optical components. With advances in semiconductor fabrication techniques, photonic integrated circuits (PICs) offer significant improvements in compactness and energy efficiency over those constructed from discrete optical components\cite{carpintero2014microwave,pang2011100,jia20202,harter2020generalized,li20191}. By integrating discrete components on a silicon-on-insulator (SOI) chip, compact, high-density, tunable, and energy-efficient photonic wireless transmitters can be realized, offering improved system stability\cite{perez2024general,cavaliere2025integrated}. PICs can be configured as tunable optical filters, such as tunable Mach-Zehnder interferometer (MZI)-based reconfigurable microring resonators that enable flexible on-chip wavelength division multiplexing (WDM) functionality\cite{liu2020integrated,zhuang2015programmable,poulopoulos2018flexible}. Most existing integrated photonic wireless transmitters incorporate only a limited number of passive devices and still depend on external active components for electro-optical conversion\cite{pan2020reflective,falconi2021reconfigurable,chen2022integrated}. Moreover, integrated implementations capable of OFC-based heterodyne generation remain rare. The integration of photonic technologies with microwave photonics creates new opportunities to combine optical, radio frequency, and digital electronic domains for processing wireless information, while enhancing both performance and energy efficiency.

In this work, we demonstrate a photonics-assisted high bandwidth wireless signal generation using a highly integrated silicon photonic chip incorporated with a reconfigurable electro-optical frequency comb. The system design of chip includes multiple tunable microring resonators (MRRs), which can be configured as on-chip filters for selective OFC tone filtering. The unique benefit of this kind of filtering element is associated with their enhanced flexibility in terms of coupling ratio and resonance wavelength. In addition, a Mach-Zehnder modulators (MZMs) array is integrated on the same chip to enable multi-channel high-speed signal modulation. The chip is capable of generating signals in different frequency bands simultaneously or synthesizing vector-modulated signals. Based on this PIC, we experimentally demonstrate generation of single-band mm-wave signals, as well as simultaneous generation of 12-Gbaud BPSK/4ASK signals with carrier frequencies of 39 GHz and 52 GHz. Furthermore, we also demonstrate the synthesis of vector-modulated mm-wave signals for arbitrary IQ signals construction, including QPSK and 16QAM formats. The proposed scheme provides a stable, low-loss, and highly integrated solution for millimeter-wave bands or terahertz signals generation and RoF applications, thereby significantly reducing the footprint and power consumption of systems.

\section{Experiments and Results}

\begin{figure}
    \centering
    \includegraphics[width=1\linewidth, height=0.376\textheight]{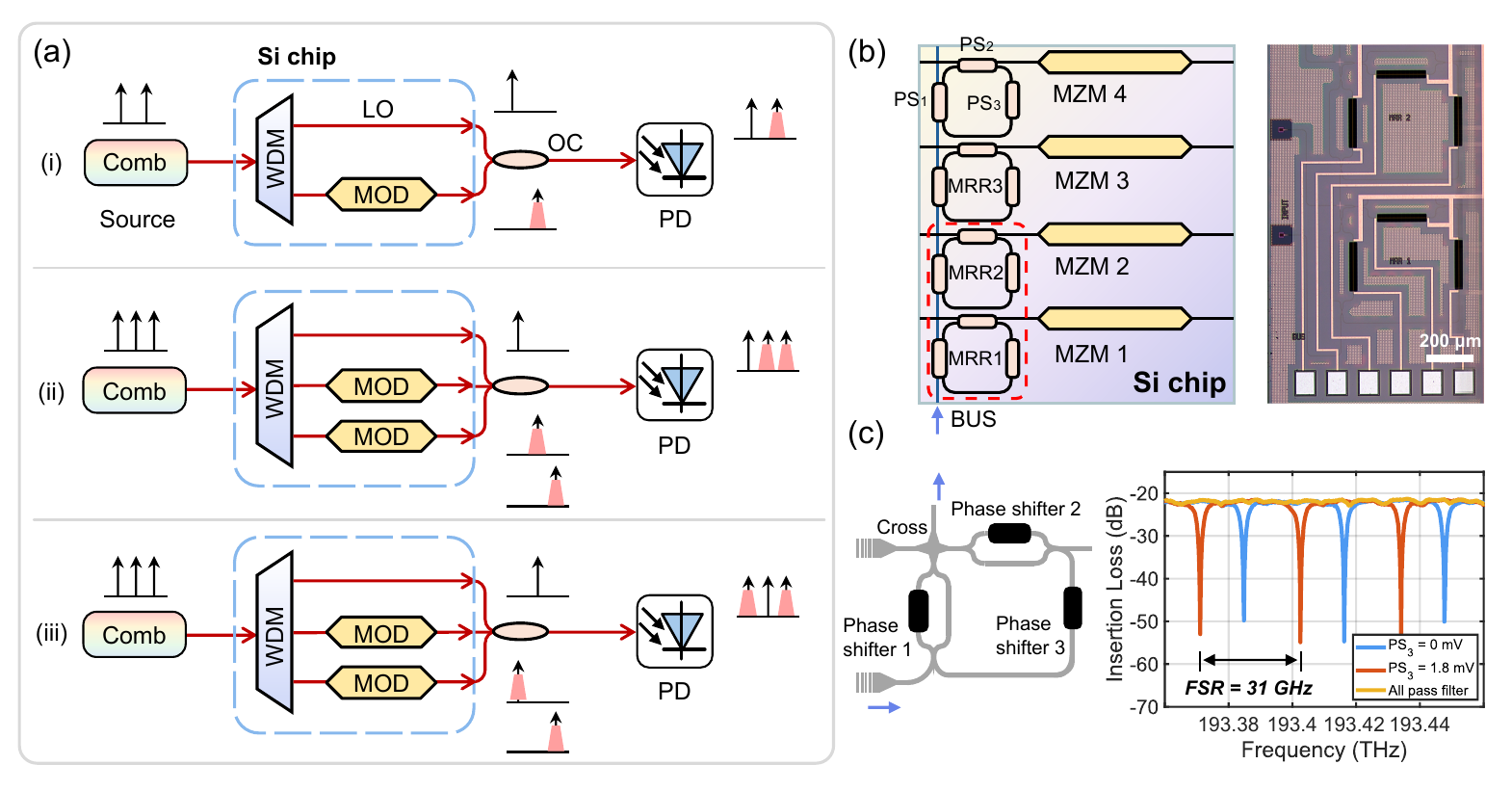}
    \caption{Schematic diagram of high-bandwidth signal generation using an integrated photonic system. (a). Experimental schematics of three compared approaches. (b). System architecture of the silicon photonic chip, with a micrograph of the reconfigurable MRRs. (c). schematics of the on-chip reconfigurable MRR and corresponding measured spectral response.}
    \label{fig:chip}
\end{figure}

We proposed three experimental configurations, as illustrated in Fig\ref{fig:chip} (a). To mitigate frequency offset and phase noise between optical carriers from two different lasers, a tunable OFC was employed as the light source, enhancing the stability of the generated signals. The repetition rate of the OFC can be flexibly adjusted by tuning the frequency of the RF source that drives the external modulator. The comb lines are separated by an on-chip wavelength-division multiplexing (WDM) module. One selected optical carrier is left unmodulated to serve as the LO, while the remaining carriers are modulated with baseband data via on-chip modulators. Scheme (i) in Fig\ref{fig:chip} (a) represents the baseline approach, producing a single-frequency signal. Scheme (ii) demonstrates simultaneous generation of multi-frequency (multi-band) signals, and scheme (iii) illustrates the path for vector signal generation. 

The components enclosed within the blue dashed line correspond to the silicon photonic integrated chip, with its detailed architecture shown in Fig\ref{fig:chip} (b). The on-chip system comprises two main sections: a reconfigurable MRR array serving as on-chip filters, and a modulator array. The MRRs are cascaded via a bus waveguide. When the input port of the bus is used as the microring input, the drop port is connected to a modulator, while the through port serves as the bus output. The measured spectral response of a microring is shown in Fig\ref{fig:chip} (c), indicating a free spectral range (FSR) of 31 GHz and extinction ratio (ER) over 30 dB. The resonators can also be configured as all-pass filters. By adjusting the electrical power applied to the thermal phase shifters on the microrings, the spectral response can be shifted to achieve precise wavelength alignment during filtering. The integrated photonic chip was fabricated on 8$''$ wafer using a commercial 220 nm SOI platform, with a 3 $\mu$m thick BOX. The on-chip modulators are based on a standard carrier-depletion scheme with a traveling-wave single-signal (SS) differential electrode design. The optical interfaces of the chip employ a fiber array (FA) for optical signal coupling, while the electrical packaging utilizes wire bonding to a printed circuit board (PCB). A four-channel differential RF driver amplifier chip is co-packaged, achieving measured modulators' bandwidth exceeding 35 GHz.

\subsection{Single frequency signal generation}
\begin{figure}
    \centering
    \includegraphics[width=1\linewidth, height=0.374\textheight]{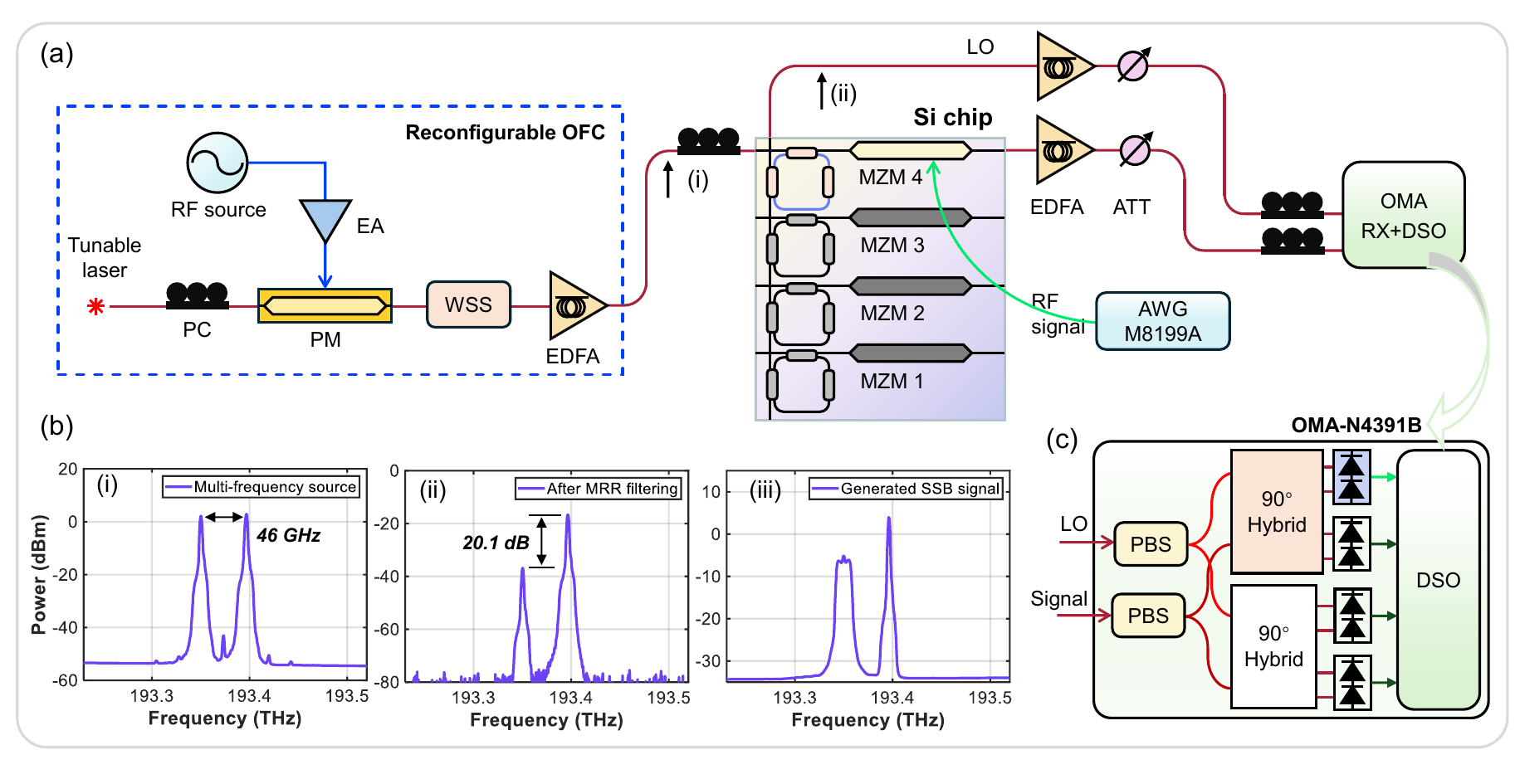}
    \caption{Schematic of single frequency signal generation. (a). Experiment setup and optical link. PC, Polarization Controller; EA, Electrical Amplifier; PM, Phase Modulator; WSS, Wavelength Selective Switch; EDFA, Erbium Doped Fiber Amplifier; ATT, Attenuator; PBS, Polarization Beam Splitter; DSO: Digital Storage Oscilloscope. (b). Measured optical spectra of the (i) multi-wavelength sources, (ii) after MRR filtering and (iii) generated high bandwidth single sideband signal, with corresponding measurement points annotated in (a). (c). Schematic diagram of internal structure of the OMA.}
    \label{fig:exp1}
\end{figure}

Figure\ref{fig:exp1} illustrates the experimental setup of Scheme 1. The reconfigurable OFC, enclosed within the blue dashed box, is generated using a commercial lithium niobate phase modulator. The resulting multi-wavelength output is filtered by a wavelength selective switch (WSS), retaining only two wavelengths as shown in Fig.\ref{fig:exp1} (b)(i). With the microwave source set at 23 GHz, the generated OFC exhibits a 46 GHz spacing between adjacent odd-order comb lines. Given the fixed 31 GHz free spectral range (FSR) of the on-chip MRR, the two filtered subcarriers are theoretically aligned at the peak and trough of the MRR spectral response, respectively. This configuration corresponds to the maximum extinction ratio condition of the MRR, ensuring optimal filtering performance.

The two tones of OFC are amplified by an erbium-doped fiber amplifier (EDFA) and coupled to MRR4 on the chip. In this configuration, MZM4 is connected to the through port of MRR4. An arbitrary waveform generator (AWG) drives the modulator with the transmitted signal. The bus waveguide output, corresponding to the drop port of MRR4, provides the optical LO signal, whose spectrum is shown in Fig\ref{fig:exp1} (b)(ii). Both the modulated signal and the LO are fed into an optical modulation analyzer (OMA), which integrates the functions of the optical coupler and photodiode depicted in the schematic. Figure\ref{fig:exp1} (c) details the internal structure of the OMA. The two input optical signals are polarization-split and mixed in a 90° hybrid, followed by heterodyne beating in a balanced photodetector (BPD) to generate the high-bandwidth electrical signal. Although the OMA supports dual-polarized IQ signal reception, here only one BPD channel is utilized. With the help of two polarization controllers (PCs), the experimental setup is similar to the structure in Fig\ref{fig:chip} (a).The generated signal is directly sampled by a high-speed oscilloscope with a bandwidth of 62 GHz at 256 GSa/s for subsequent offline DSP. The optical spectrum after combining the modulated signal and LO via an optical coupler is shown in Fig\ref{fig:exp1} (b)(iii), confirming the generation of a single-sideband signal.

\begin{figure}
    \centering
    \includegraphics[width=0.7\linewidth, height=0.385\textheight]{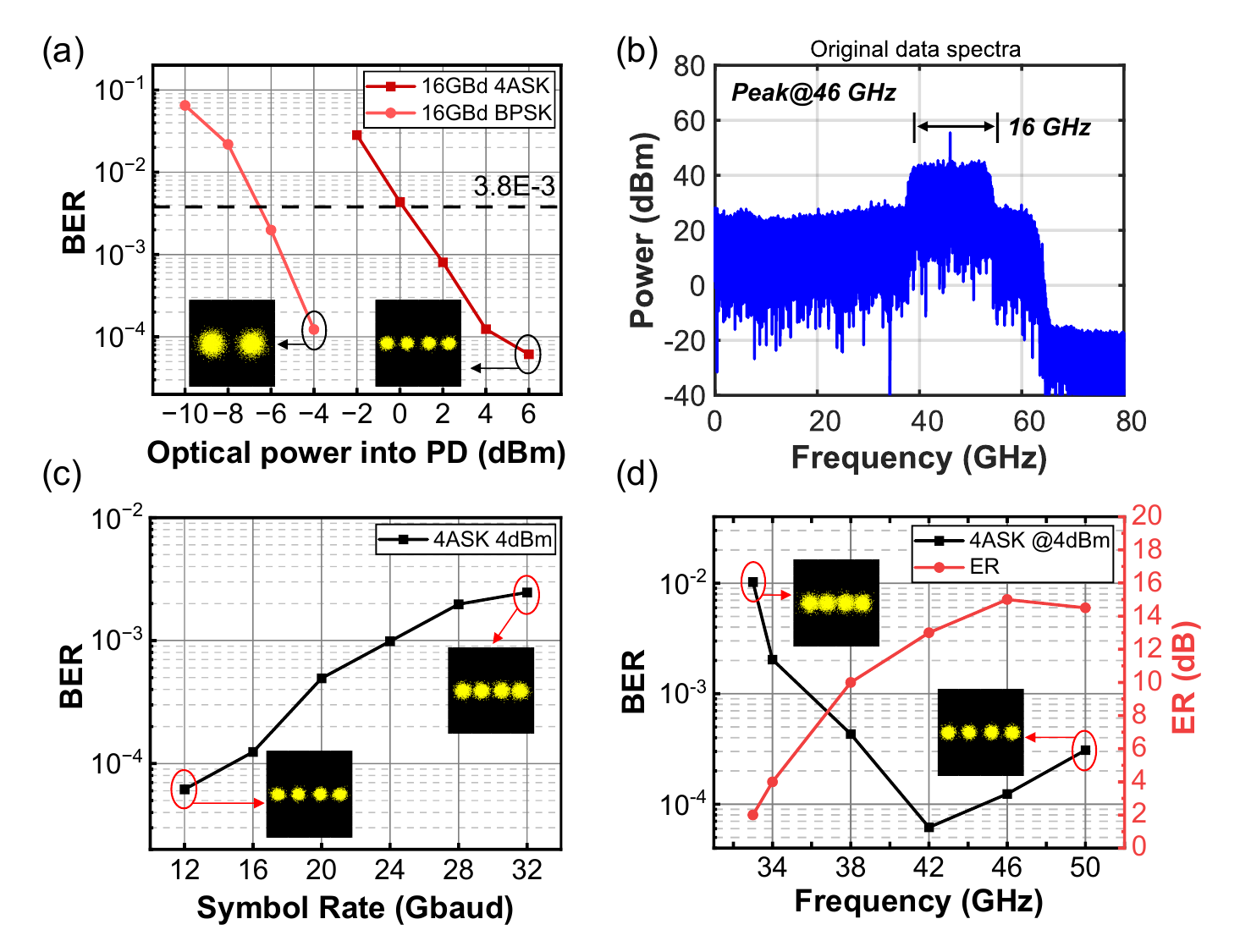}
    \caption{Experimental results of single frequency signal generation. (a). Measured BER versus optical power into PD curve for modulated 16 Gbaud BPSK and 16 Gbaud 4ASK signals. Insert is corresponding constellation diagrams. (b) Electrical spectrum of the captured 16 Gbaud BPSK modulated signal. (c) BER performance versus modulated signals' symbol rate. (d) BER versus generated signals' frequency.}
    \label{fig:ex1results}
\end{figure}

We first characterized the bit error rate (BER) as a function of the optical power into the PD at Rx, as shown in Fig\ref{fig:ex1results} (a). For a generated bandwidth of 46 GHz signal modulated with 16 Gbaud BPSK, the receiver sensitivity at the hard decision forward error correction (HD-FEC) threshold was measured to be -7 dBm. In comparison, the sensitivity for a 16 Gbaud 4ASK signal was 0 dBm. The electrical spectrum of the captured signal is presented in Fig.\ref{fig:ex1results} (b), showing a spectral peak at 46 GHz corresponding to the generated carrier frequency. With a received optical power of 4 dBm, we evaluated the BER performance across different symbol rates, as summarized in Fig\ref{fig:ex1results} (c). The results indicate a progressive increase in BER with higher symbol rates. Furthermore, the BER dependence on the generated carrier frequency was investigated. Owing to the fixed 31 GHz FSR of the on-chip reconfigurable MRRs, the maximum ER at which consequently the optimal filtering performance was achieved for two optical carriers separated by 46 GHz, as evidenced by the red curve in the Fig\ref{fig:ex1results} (d). The measured BER at an input PD optical power of 4 dBm, show that the BER progressively increased as the signal frequency decreased below 42 GHz. Theoretically, the minimum BER was expected at a generated frequency of 46 GHz. However, elevated receiver noise at higher frequencies resulted in increased BER for signals with carrier frequencies above this optimal value.

\subsection{Multi-frequency signal generation}

\begin{figure}
    \centering
    \includegraphics[width=1\linewidth, height=0.288\textheight]{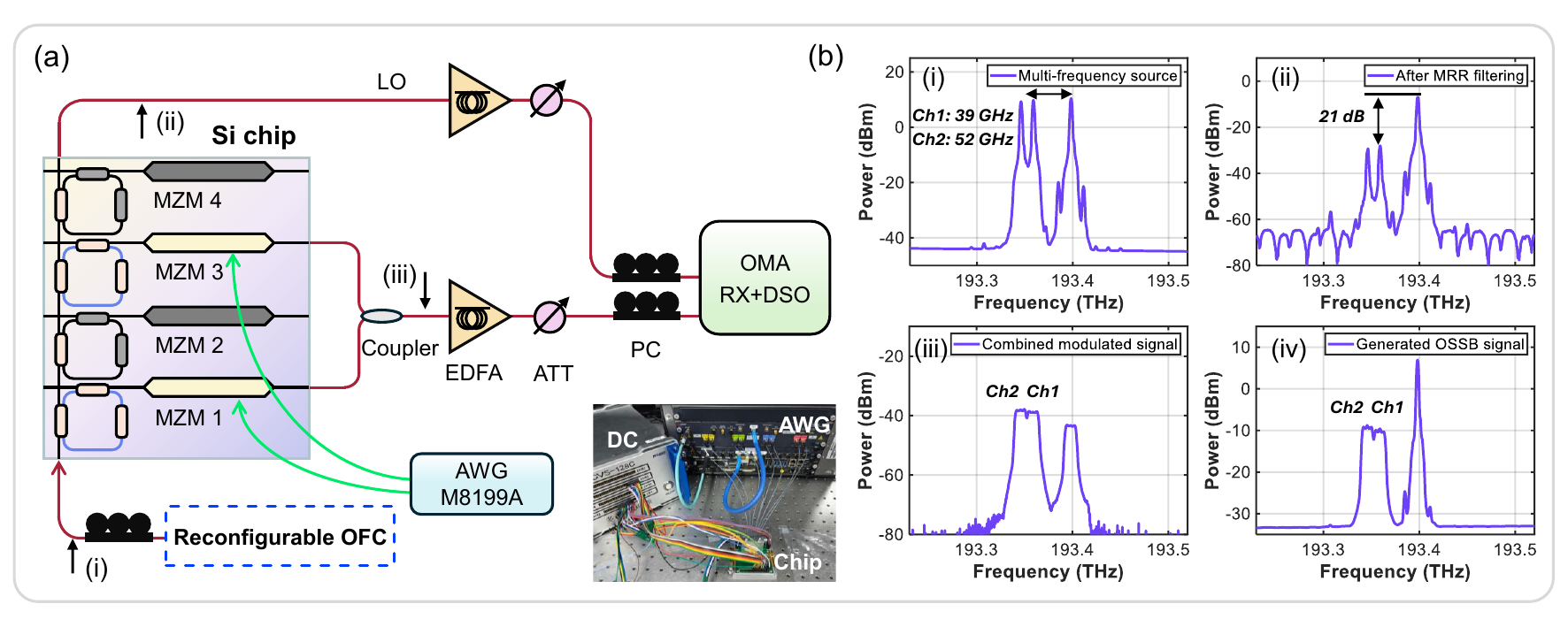}
    \caption{Schematic of multi-frequency signal generation. (a). Experiment setup and optical link. The inset shows a photograph of the packaged photonic integrated chip deployed in the experimental system: the 128-channel programmable voltage source is positioned on the left, and the arbitrary waveform generator is located at the top. (b). Measured optical spectra of the (i) multi-wavelength sources, (ii) after MRR filtering, (iii) combined modulated signal and (iv) generated high bandwidth multi-frequency single sideband signals, with corresponding measurement points annotated in (a).}
    \label{fig:exp2}
\end{figure}

We further demonstrate simultaneous generation of multi-band/multi-frequency signals, as shown in Fig.\ref{fig:exp2}. The repetition rate of the reconfigurable OFC was set to 13 GHz, resulting in three remaining optical carriers after WSS filtering, exhibiting mutual frequency spacings of 13 GHz and 39 GHz. This configuration ensures that the two closer tones maintain the same ER relative to the more distant tone-which serves as the LO-during on-chip filtering, thereby optimizing filtering performance and guaranteeing high signal quality. The optical spectrum of the multi-wavelength source is shown in Fig.\ref{fig:exp2} (b)(i), the source signal is then fed into the bus waveguide input and filtered by MRR1 and MRR3. The remaining two MRRs are configured as all-pass filters, with the bus output serving as the LO signal. As shown in Fig.\ref{fig:exp2} (b)(ii), an ER of 21 dB is achieved relative to the other two optical carriers, while spectral imperfections on both sides result from the cascaded spectral response of the two filters. The modulators in channels 1 and 3 are driven by independent signals, and the modulated outputs are combined via an optical coupler, as depicted in Fig.\ref{fig:exp2} (b)(iii). The combined modulated signal and the LO are simultaneously fed into the OMA, enabling concurrent generation of modulated signals in different frequency bands. Theoretically, this approach can be extended to generate a larger number of simultaneous single-sideband signals. However, this imposes more stringent requirements on the filters. In the case of MRR filters, not only is a higher ER necessary to achieve more complete filtering, but a broader FSR tuning range is also required to accommodate wavelength-specific adjustments.

We characterized the multi-frequency signal generation performance by measuring the BER as a function of incident optical power in the PD, as shown in Fig.\ref{fig:ex1results} (a). For simultaneous transmission of two 12 Gbaud BPSK signals at different frequencies, the receiver sensitivity at the HD-FEC threshold was measured at -3 dBm for Channel 1 (Ch1) and -5 dBm for Channel 2 (Ch2), representing a 2 dB difference. The combined modulated signals exhibit a 2-3 dB optical power disparity between Ch1 and Ch2, attributable to non-ideal microring filtering performance and non-uniformity among on-chip modulators. This discrepancy is also subtly visible in the corresponding optical spectrum. Consequently, Ch1 experiences degraded signal quality, resulting in a higher BER curve compared to Ch2. A similar trend was observed for 12 Gbaud 4ASK modulation. At the 20$\%$ soft decision forward error correction (SD-FEC) threshold, the receiver sensitivity was 3 dBm for Ch1 and 0 dBm for Ch2, corresponding to a 3 dB penalty. The superior signal quality of Ch2 under identical received optical power conditions is further corroborated by the constellation diagrams in Fig.\ref{fig:ex1results} (b). 

\begin{figure}
    \centering
    \includegraphics[width=0.7\linewidth, height=0.234\textheight]{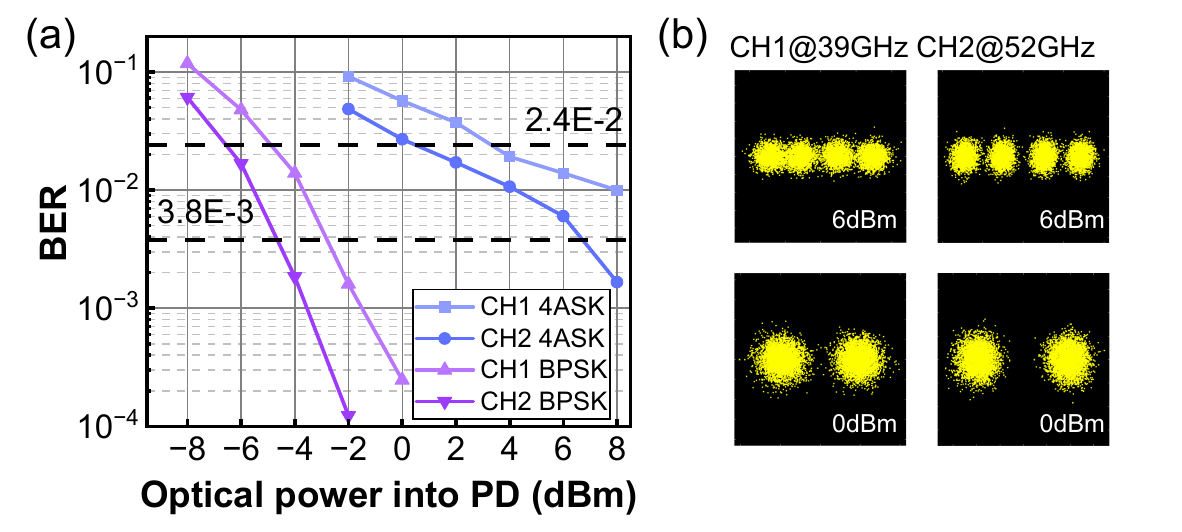}
    \caption{Experimental results of multi-frequency signal generation. (a). Measured BER versus optical power into PD curve for modulated 12 Gbaud BPSK and 12 Gbaud 4ASK signals. (b). Corresponding constellation diagrams of generated signals.}
    \label{fig:ex2results}
\end{figure}

\subsection{Vector signal generation}

\begin{figure}
    \centering
    \includegraphics[width=0.7\linewidth, height=0.365\textheight]{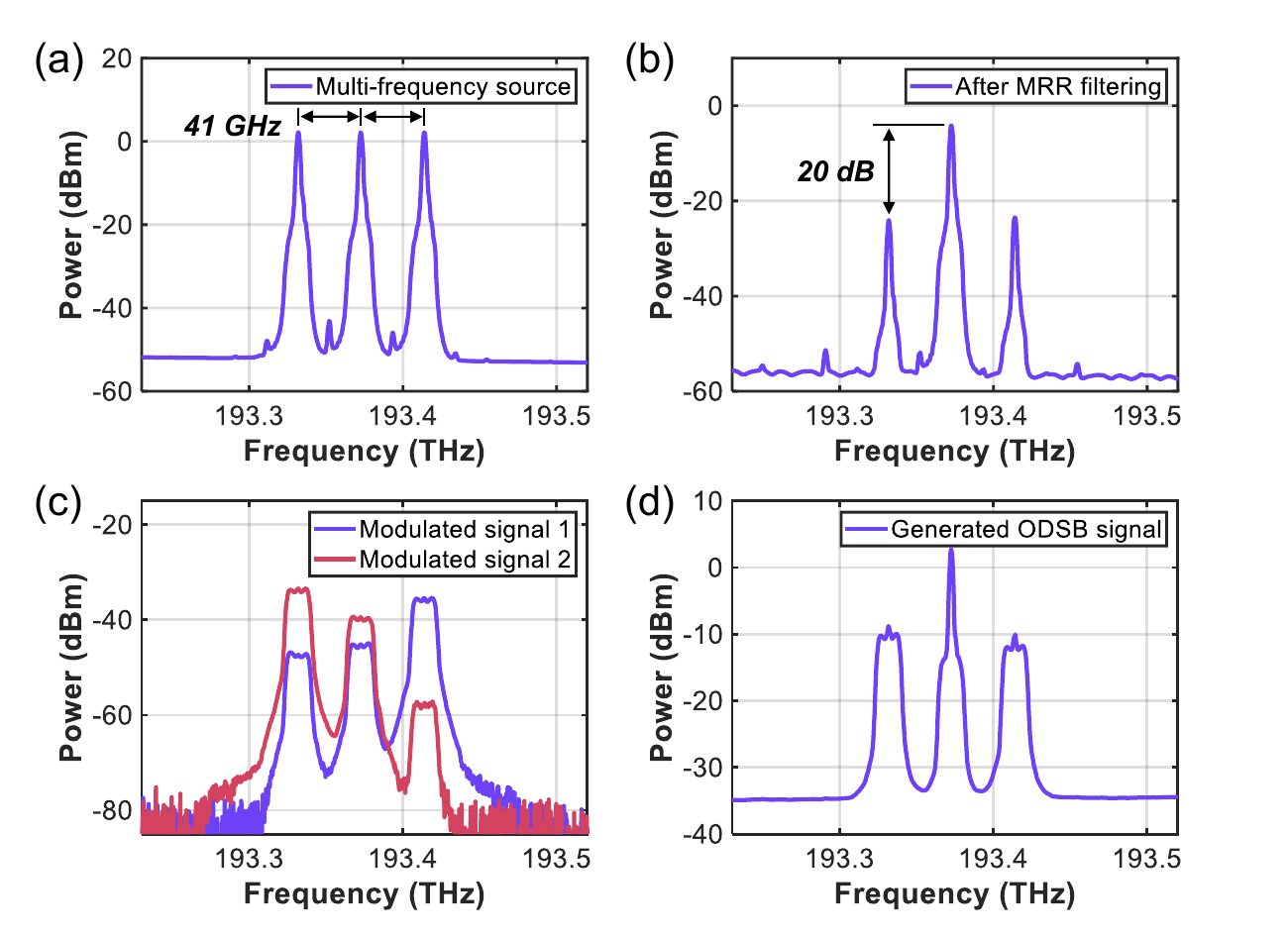}
    \caption{Measured optical spectra of generation of vector signals. (a). Multi-frequency light source. (b). Central subcarrier after MRR filtering serves as the local oscillator. (c). Modulated signals of two subcarriers. (d). Generated optical double sideband signal. }
    \label{fig:exp3}
\end{figure}

\begin{figure}
    \centering
    \includegraphics[width=0.7\linewidth, height=0.22\textheight]{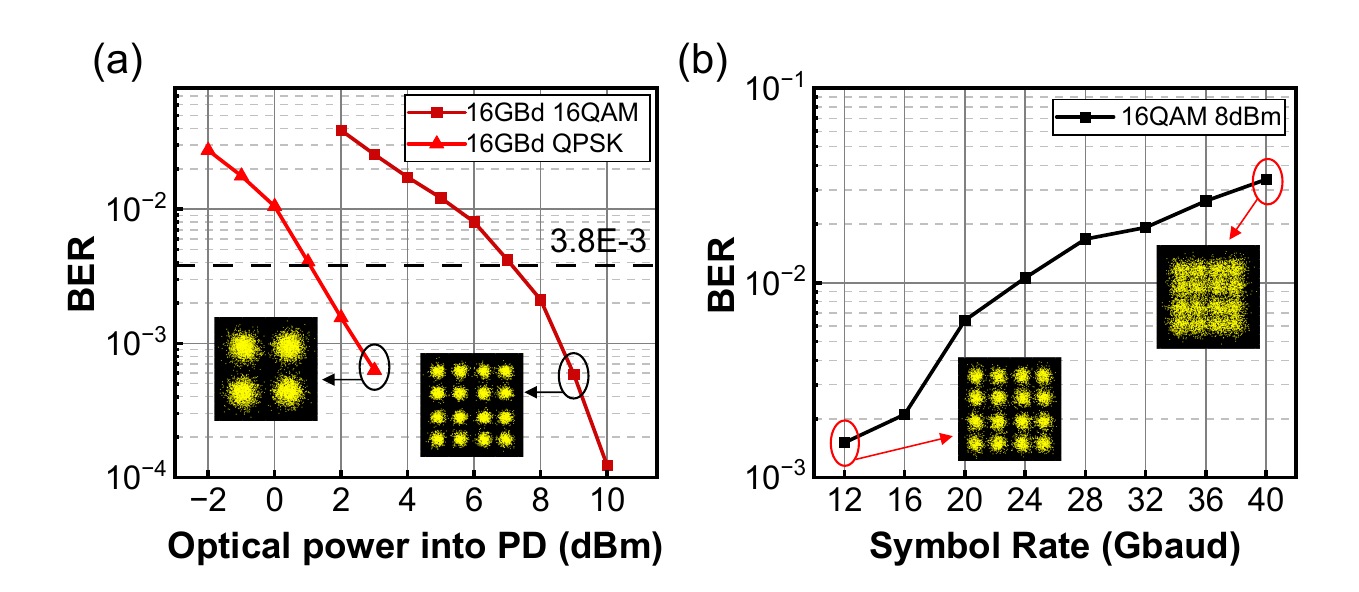}
    \caption{Experimental results of vector signals generation. (a). Measured BER versus optical power into PD curve for modulated 16 Gbaud QPSK and 16 Gbaud 16QAM signals. Insert is corresponding constellation diagrams. (b). BER performance versus generated 16 QAM signal's symbol rate.}
    \label{fig:ex3results}
\end{figure}

Finally, we demonstrate the synthesis of vector signals using an experimental setup consistent with Fig.\ref{fig:exp2}, with modifications to the multi-frequency optical source configuration. As shown in Fig.\ref{fig:exp3}, three optical carriers with uniform frequency spacing of 41 GHz are employed. The central subcarrier serves as the LO. After on-chip filtering, it achieves a 20 dB ER relative to the two outer carriers. The remaining two optical carriers are modulated with BPSK and 4ASK data streams, respectively, functioning as the in-phase (I) and quadrature (Q) components of the final synthesized vector signal, as illustrated in Fig.\ref{fig:exp3} (c). The two modulated signals and the LO undergo heterodyne beating by the PD, generating vector mm-wave signals at 41 GHz in QPSK and 16QAM formats. The measured BER performance of these synthesized vector signals is shown in Fig.\ref{fig:ex3results} (a). At the HD-FEC threshold, the receiver sensitivity is 1 dBm for QPSK and 7 dBm for 16QAM. Figure\ref{fig:ex3results} (b) further demonstrates that higher symbol rates lead to increased BER at identical received optical power levels, a trend also evident in the corresponding constellation diagrams. The proposed scheme is equal to single frequency millimeter-wave signal generation with IQ modulation formats such as QPSK and QAM, although the on-chip modulator configuration in this work was demonstrated using BPSK and 4ASK modulation.

\section{Conclusion}

The design of on-chip MRR can significantly influence the experimental results. The performance of optical filters depends on multiple parameters, among which the Q-factor, extinction ratio and free spectral range are of great importance\cite{xu2025progress}. The ER is strongly correlated with the waveguide propagation loss: lower propagation loss leads to a higher Q-factor and consequently a larger ER. A high ER significantly improves filtering selectivity and reduces crosstalk caused by residual leaked wavelength components in the combined signal. Strategies for enhancing the ER include employing cascaded or high-order resonator architectures, or implementing post-fabrication tuning techniques for performance correction\cite{little2000filter,long2014optically}. Furthermore, the reconfigurability of optical filters-particularly the tunability of bandwidth and FSR-is crucial for enabling dynamic channel allocation, thereby improving spectral utilization efficiency and overall transmission capacity. In our experimental context, adjusting the repetition rate of the OFC relative to the FSR is essential for achieving optimal filtering performance. The FSR of an MRR can be dynamically tuned by modifying the effective cavity length or reducing the length of the MZI arms\cite{fu2024optical,bogaerts2012silicon}.

In this paper, we demonstrate photonics-assisted mm-wave signal generation using a photonic integrated chip and reconfigurable OFC, overcoming the bandwidth limitations of electronic components. The OFC source eliminates phase drift between the local oscillator and modulated optical signals while enabling flexible frequency tuning through repetition rate adjustment. The silicon photonic integrated chip implements on-chip filtering and signal modulation through reconfigurable microring resonators, offering significant advantages in footprint and power efficiency over conventional discrete photonic devices. We experimentally demonstrate the generation of a single-frequency 46 GHz signal modulated with 16 Gbaud BPSK and 4ASK formats, evaluating the BER performance and investigating the impact of microring characteristics on signal quality. The system's capability for simultaneous multi-frequency signal generation is also verified through the successful transmission and reception of 39 GHz and 52 GHz signals. Furthermore, we synthesize and characterize vector signals by generating 16 Gbaud QPSK and 16QAM formats with corresponding BER measurements. Given the periodic nature and broad spectral coverage of the microring resonator's FSR across the entire C-band, the chip is expected to generate signals in higher frequency bands-extending from millimeter-wave to terahertz regimes-when integrated with high-bandwidth photodetectors such as uni-traveling-carrier photodiodes (UTC-PDs). These experimental results validate the flexible configurability of the proposed approach, providing new insights for photonics-assisted mm-wave signal generation in applications such as radio-over-fiber systems.

\section*{Acknowledgments}
This work was supported by The Major Key Project of PCL, National Talent Program, The Guangdong Basic and Applied Basic Research Foundation (2024A1515030297), National Natural Science Foundation of China (62105209). 

%Bibliography
\bibliographystyle{unsrt}  
\bibliography{references}

\end{document}